\documentclass[aps, prl, reprint, showpacs, twocolumn, 10pt, groupedaddress]{revtex4-1}

\usepackage{graphicx, amsmath, dsfont, dcolumn, bm, times, color, braket, amssymb}

\begin{document}

\bibliographystyle{apsrev4-1}

\title{Fine energy splitting of overlapping Andreev bound states in multi-terminal superconducting nanostructures}

\author{Viktoriia Kornich}
\affiliation{Kavli Institute of Nanoscience, Delft University of Technology, 2628 CJ Delft, The Netherlands}
\author{Hristo S. Barakov}
\affiliation{Kavli Institute of Nanoscience, Delft University of Technology, 2628 CJ Delft, The Netherlands}
\author{Yuli V. Nazarov}
\affiliation{Kavli Institute of Nanoscience, Delft University of Technology, 2628 CJ Delft, The Netherlands}

\date{\today}

\begin{abstract}
The recent proposals of experiments with single Andreev bound states make relevant a detailed analysis of these states in multi-terminal superconducting nanostructures. We evaluate the energy splitting of degenerate Andreev bound states, that overlap in a superconducting lead, and find that the splitting is reduced in comparison with their energy by a small factor $\sqrt{R G_Q}$, $R G_Q$ being the dimensionless resistance of the overlap region in the normal state. This permits quantum manipulation of the quasiparticles in these states. We provide a simple scheme of such manipulation.     
\end{abstract}

\maketitle

\let\oldvec\vec
\renewcommand{\vec}[1]{\ensuremath{\boldsymbol{#1}}}
The superconducting nanodevices are in focus of modern experimental research, in particular because they are a promising platform for various qubit realizations, e.g. Josephson-based qubits of several kinds \cite{nakamura:prl97, han:science01, vion:science02, martinis:prl02, berkley:science03} or Majorana bound states \cite{kitaev:physusp01, lutchyn:prl10, oreg:prl10, alicea:rpp12, mourik:science12}. These structures, containing superconductor-semiconductor or superconductor-insulator junctions, host Andreev bound states (ABS), which can also be used as a qubit \cite{chtchelkatchev:prl03, janvier:science15}. Andreev reflection between normal metal and superconductor was first discussed in Ref.~\cite{andreev:jetp64}, and has been a subject of intense theoretical and experimental research \cite{blonder:prb82, shelankov:ftt84, soulen:science98, bergeret:prl10, deacon:prl10, chiodi:scirep11, lee:nnano13, bretheau:prb14, bretheau:natphys17, tosi:prx19} that spans far beyond quantum information topics.

It is known and commonly used that the ABS forming in nanostructures are defined by the properties of the nanostructure, not depending on the details of electron scattering in the adjacent superconducting leads, which is a consequence of Anderson's theorem \cite{anderson:jpcs59}. For short nanostructures between two leads, each transport channel with transmission $T$ gives an ABS at the energy \cite{beenakker:prl91} $E=\pm\Delta\sqrt{1-T\sin^2{\left[(\phi_1-\phi_2)/2\right]}}$, $\Delta$ being the superconducting gap in the leads, $\phi_{1,2}$ being the superconducting phases of the leads. ABS extend to the leads at distances of the order of the correlation length $\xi_0$, this is much larger than the nanostructure size. ABS can be realized in semiconducting-nanowire-superconducting structures, the technology of those has advanced strongly owing to the applications in the field of Majorana bound states \cite{mourik:science12, albrecht:nature16, vanwoerkom:natphys17, goffman:njp17, hays:prl18}, and can be characterized experimentally \cite{vanwoerkom:natphys17, goffman:njp17}.
There is much recent progress in multi-terminal devices \cite{plissard:nnano13, pankratova:arxiv18, draelos:nanolett19} that has been partially inspired by theoretical predictions of Weyl points and quantized transconductance \cite{riwar:natcom16}.% Much is known about single ABS at the junction, however the interaction of ABS might bring new phenomena as well as new technologies and devices.

\begin{figure}[b]
\begin{center}
\includegraphics[width=0.85\linewidth]{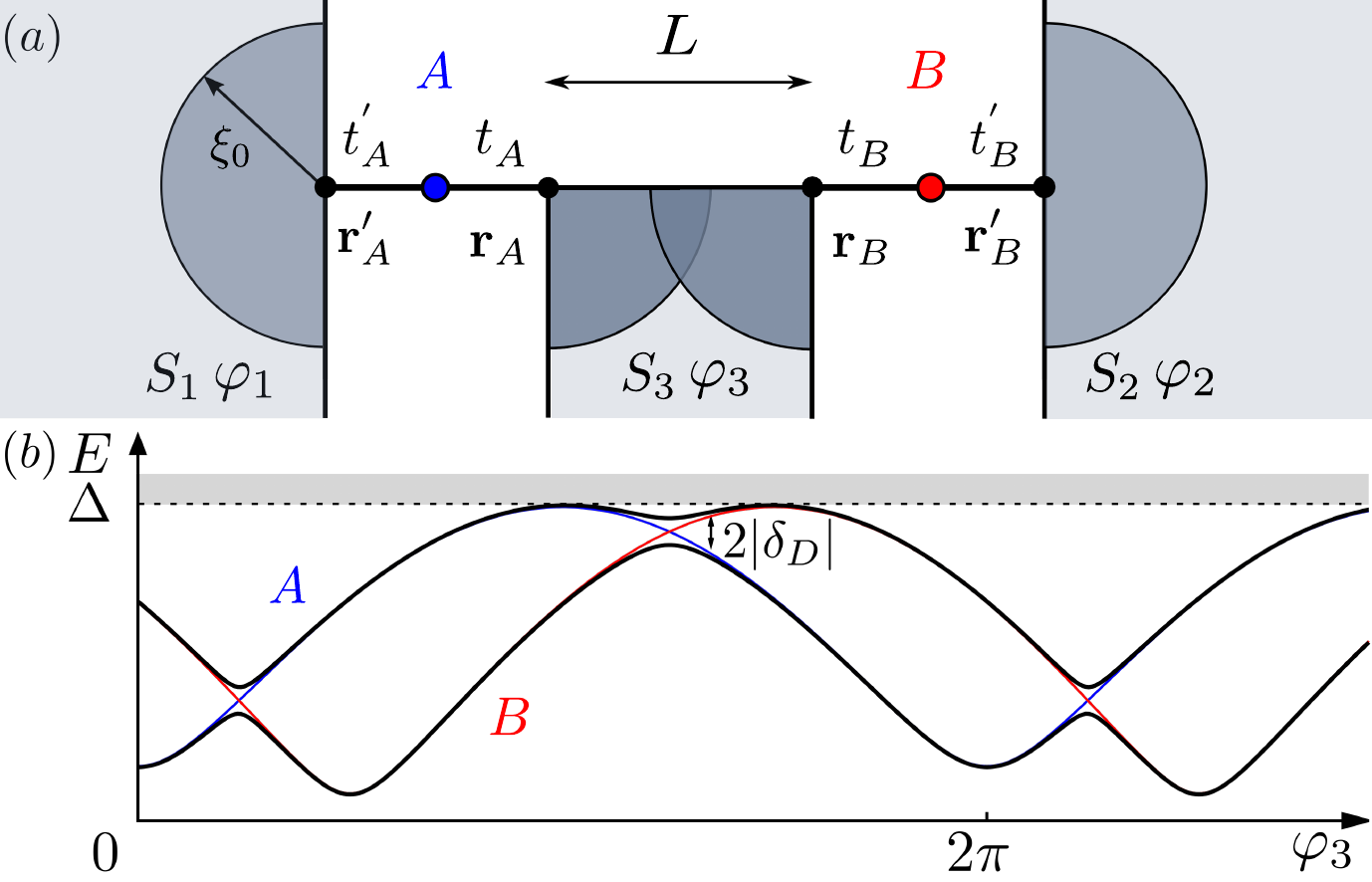}
\caption{(a) The "Andreev molecule" setup. Two ABS are formed in the two single-channel junctions, A and B, that are separated by $L$ and connect three superconducting leads $S_{1,2,3}$ with the same $\Delta$. The ABS wave functions spread over the scale of $\xi_{0}$, overlapping in $S_3$. (b) The energy spectrum of ABSs versus $\varphi_{3}$ ($\varphi_{1} = \pi$, $\varphi_{2} = 3\pi/2$, the junction transmission coefficients being $T_{A}=0.95$, $T_{B}=0.98$), manifests avoided crossings at the degeneracy points. We show that the energy splitting at the crossings is fine even for a significant overlap, $\delta_D \propto \Delta \sqrt{R G_{Q}}$, where $R$ being normal resistance of the overlap region.}
\label{fig:Andreev_molecule}
\end{center}
\end{figure}

Very recent experimental and theoretical developments concern so-called Andreev molecules in various layouts \cite{su:natcom17, pillet:arxiv18, scheruebl:bjn19}. The term ``molecule'' refers to the situation where two single ABS have close energies, this enables their hybridization and formation of the superpositions. Refs.~\cite{su:natcom17, scheruebl:bjn19} discuss ABS in quantum dots, where ABS overlap through the tunnel barrier separating the dots. An interesting alternative has been put forward in Ref.~\cite{pillet:arxiv18}. The proposed three-terminal setup encompasses two single-channel junctions that connect three superconducting leads, see Fig.~\ref{fig:Andreev_molecule} (a). Two single ABS may overlap and hybridize in the middle lead. The overlap and the corresponding energy splitting must cease exponentially as $\exp{(-L/\xi_0)}$, provided the separation of the junctions $L\gg \xi_0$. The authors of Ref.~\cite{pillet:arxiv18} indicate that Andreev molecules have potential applications in quantum information, metrology, sensing, and molecular simulation. 

In this Letter we have evaluated the energy splitting of overlapping ABS and have comprehended it as an interference effect similar to mesoscopic fluctuations of conductance \cite{chakravarty:pr86}, which develops in the lead on the scale of ABS overlap that encompasses a large number of quantum channels. We have estimated the typical energy splitting $\delta_D\sim \Delta/\sqrt{N}$, $N$ being the number of channels, that can be estimated as the inverse of the normal dimensionless resistance of the overlap region, $N\approx (RG_Q)^{-1}$, $G_Q\equiv e^2/(\pi\hbar)$ being conductance quantum. Therefore $\delta_D$ remains fine at the energy scale of $\Delta$. This facilitates quantum manipulations, an example of which we give. We have derived concrete expressions for $\langle |\delta_D|^2\rangle$, relating it to semiclassical propagation of an electron between the junctions, and employed it for an experimentally relevant setup. Observation of energy splitting gives an interesting and unusual way to see and explore mesoscopic fluctuations.

Let us first describe the setup under consideration (Fig.~\ref{fig:Andreev_molecule}) in general terms specifying the details later. The setup consists of three superconducting leads, connected by two single-channel junctions, and there is an ABS formed in each junction. If one neglects their hybridization, their energies are $E_{A(B)}=\Delta\sqrt{1-T_{A(B)}\sin^2{\left[(\phi_{1(2)}-\phi_3)/2\right]}}$. We plot the energies in Fig.~\ref{fig:Andreev_molecule} (b) for $\phi_1-\phi_2=\pi/2$ and $T_A=0.95$, $T_B=0.98$; the degeneracy at $E_A=E_B$ is avoided. The separation between the junctions is $L\gg \lambda_F$, $\lambda_F$ being the Fermi wave length. This implies that the electron transport in the region covered by the ABS wave functions, occurs in a big number of transport channels. The exact number depends on the geometry of the device, material and morphology of the leads, and the detailed characteristics of electron transport, that can be ballistic, diffusive, or intermediate between the two. At the level of an estimate, all these details can be incorporated into the effective resistance $R$ of the region spanned by the ABS wave functions, so that $N \sim (RG_Q)^{-1}$. The wave functions of the ABS overlap as shown in Fig.~\ref{fig:Andreev_molecule} (a) and hybridize. The hybridization is big near each avoided crossing and can be described with an off-diagonal matrix element $\delta_D$, which is a complex number.  

Let us estimate the energy splitting $2|\delta_D|$ near an avoided crossing. The energies of the states are $E_{A,B}\sim\Delta$. These states are formed by electrons coming in and out of a junction to$/$from the adjacent leads and returning as holes to the same junction. The electron wave function is distributed among $N$ transport channels involved. The contributions of different channels to Andreev scattering amplitude come with the same sign and phase, this is precisely the reason for the energy of ABS not to depend on the details of the scattering in the leads. This implies that the contribution of each channel to the ABS energy can be estimated as $\Delta/N$. As to $\delta_D$, it is determined by the electron and hole propagation from the junction $A$ to the junction $B$. Since these points are distinct and separated by $L\gg \lambda_F$ one expects the contributions of different channels to come with the different and generally random complex amplitudes. These random amplitudes may be related to mesoscopic fluctuations of electron propagation between the junctions A and B. Averaging over these random amplitudes leads to vanishing $\langle \delta_D\rangle=0$. The average $\langle|\delta_D|^2\rangle$ is contributed by independent contributions of $N$ channels and therefore the energy splitting can be estimated as $|\delta_D|\sim\Delta/\sqrt{N}\sim \Delta\sqrt{RG_Q}$.  

The junctions between the superconducting leads may have various characteristics, such as disorder, shape, material. It is known however \cite{beenakker:rmp97} that the only characteristic relevant for ABS is the transmission of these junctions. Therefore we are free to choose any microscopic model so we opt for a convenient resonant impurity model \cite{cuevas:prb01, levyyeyati:prb97}, that involves a localized state of energy $E_{\rm imp}$ with the tunnel couplings $t$ and $t'$ to the leads. We label with $A$ and $B$ these characteristics in the corresponding junctions, see Fig.~\ref{fig:Andreev_molecule} (a). The width of the resonant level is given by $\Gamma=2\pi\nu(|t|^2+|t'|^2)$, $\nu$ being the density of states assumed equal in all leads. To model weak energy dependence of the scattering we set $\Gamma\gg \Delta$, so the transmission coefficient of the junction $A$ is 
\begin{equation}
T_A=\frac{4\pi^2\nu^2|t_A|^2|t_A'|^2}{(\Gamma_A/2)^2+E_{{\rm imp},A}^2},     
\end{equation}
and similar for the junction $B$.

\begin{figure*}[tb]
\begin{center}
\includegraphics[width=\linewidth]{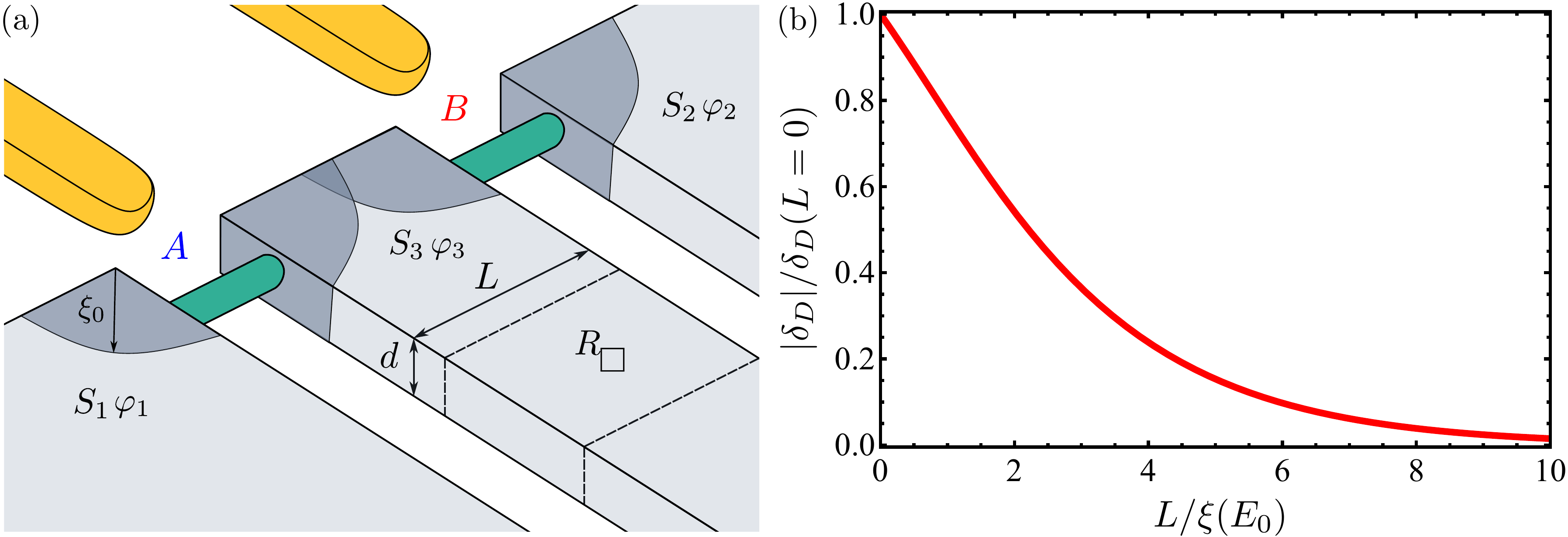}
\caption{(a) The concrete setup under consideration comprises three superconducting leads covering a single-channel semiconducting nanowire. Two hybridizing ABSs are forming at the junctions $A$, $B$. The gates (yellow) affect the potential in the wire and might be used to tune the transmission of the junctions. The middle lead is a film of thickness $d$ and width $L$ and is characterized by the resistance per square $R_{\Box}$. The ABS wave functions overlap strongly provided $L\lesssim \xi_0$. (b) Dependence of the energy splitting $|\delta_D|/\delta_D(L=0)$ on $L$. The splitting vanishes exponentially upon increasing $L$.}
\label{fig:F_L_xi}
\end{center}
\end{figure*}
To find the ABS energies we derive a Dyson equation for the Green's function $G_{ij}(E)$ defined at the resonant impurities $i,j=\{A,B\}$: $G(E)=([G^0]^{-1}-\Sigma)^{-1}$, where the blocks are the matrices in the Nambu space $G^0_{AA,BB}=(E-E_{{\rm imp},A,B}\sigma_z)^{-1}$, and the self-energy part $\Sigma$ describes the tunneling. The diagonal blocks are $\Sigma_{AA}=\mathcal{T}_{A}'G({\bm r}'_A,{\bm r}'_A)\mathcal{T}_{A}'^*+\mathcal{T}_{A}G({\bm r}_A,{\bm r}_A)\mathcal{T}_{A}^*$, where $\mathcal{T}_j=(t_j^*(\sigma_z+\sigma_0)+t_j(\sigma_z-\sigma_0))/2$, $j=\{A,B\}$, and $G({\bm r},{\bm r}')$ is a superconducting Green's function in the corresponding leads upon neglecting the tunneling, $\Sigma_{BB}$ is similar. The Green's function in close points does not depend on the details of the scattering in the lead, this is a consequence of Anderson's theorem \cite{anderson:jpcs59}. The non-diagonal blocks $\Sigma_{AB}$ and $\Sigma_{BA}$ are  $\Sigma_{AB}=\mathcal{T}_AG({\bm r}_A,{\bm r}_B)\mathcal{T}_B^*$ and $\Sigma_{BA}=\mathcal{T}_BG({\bm r}_B,{\bm r}_A)\mathcal{T}_A^*$. We see that the diagonal blocks contain Green's functions in coinciding points, while non-diagonal ones contain Green's functions in two points separated by $L\gg \lambda_F$. Since Green's functions are associated with propagation amplitudes $G({\bm r}_A,{\bm r}_B)\ll G({\bm r}_A,{\bm r}_A),G({\bm r}_B,{\bm r}_B)$. Thus $\Sigma_{AB}, \Sigma_{BA}\ll \Sigma_{AA}, \Sigma_{BB}$, and can be handled by means of the degenerate perturbation theory. This already implies $|\delta_D|\ll \Delta$.

The energies of ABS correspond to the poles of $G(E)$ \cite{altland:book10}, so we need to find zero eigenvalues of $G^{-1}(E)$. We find zero eigenvalues in each diagonal block and project $G^{-1}(E)$ onto the corresponding eigenvectors $|A\rangle$ and $|B\rangle$. We work near the crossing point $E_0$ where the unperturbed ABS energies are almost degenerate $E_A\approx E_B\approx E_0$. Expanding up to linear order near the crossing point and transforming $G^{-1}(E)$ we find that ABS energies are obtained from the effective Hamiltonian describing the level repulsion, $H_{\rm eff}=\begin{pmatrix}
E_A & \delta_D\\
\delta_D^* & E_B
\end{pmatrix}$, where  
\begin{equation}
\label{eq:delta}
\delta_D=-\frac{\langle A|\Sigma_{AB}|B\rangle\sqrt{\Delta^2-E_0^2}}{\sqrt{\Gamma_A\Gamma_B}}\propto G({\bm r}_A,{\bm r}_B). 
\end{equation}
The Green's function $G({\bm r}_A,{\bm r}_B)$ changes much on a scale of $\lambda_F$ upon changing the position of ${\bm r}_B$. This is the origin of mesoscopic fluctuations in electron transport \cite{nazarov:book09}. The components of $G({\bm r}_A,{\bm r}_B)$ can be regarded as random values with zero averages. The informative quantities are the products of these components averaged over ${\bm r}_B$ at the scale exceeding $\lambda_F$. These averaged products can be expressed with a normal-state quasiclassical propagator \cite{hekking:prb94} $\mathcal{P}({\bm r}_A,{\bm r}_B, t)$, that gives the probability to find an electron at ${\bm r}_B$ at the time moment $t$ provided it was at ${\bm r}_A$ at $t=0$ (Greek letters denote Nambu indices): 
\begin{eqnarray}
\label{eq:GG}
\nonumber
&&\langle G({\bm r}_A,{\bm r}_B)^{\alpha\gamma*}G({\bm r}_A,{\bm r}_B)^{\beta\nu}\rangle=\\ \nonumber&&=\frac{\pi^2\nu}{\Delta^2-E^2}\int dt \mathcal{P}({\bm r}_A,{\bm r}_B,t) e^{-2\sqrt{\Delta^2-E^2}|t|}\\ \nonumber&&\times[\delta_{\alpha \gamma}\delta_{\beta \nu}\delta_{\alpha \beta}(2\Delta^{2}-E^{2})\\ \nonumber
	&&+\Delta E ((1-\delta_{\alpha \gamma})\delta_{\beta \nu}e^{-i\epsilon_{\alpha \gamma}\varphi_{3}}-\delta_{\alpha \gamma}(1-\delta_{\beta \nu})e^{i\epsilon_{\beta \nu}\varphi_{3}})\\ \nonumber
	&&+\Delta^{2}(\delta_{\alpha \beta}\delta_{\gamma \nu}(1-\delta_{\alpha \gamma})-\delta_{\alpha \gamma}\delta_{\beta \nu}(1-\delta_{\alpha \beta})\\ &&+\delta_{\alpha \nu}\delta_{\gamma \beta}(1-\delta_{\alpha \gamma}) e^{-i 2 \epsilon_{\alpha \gamma}\varphi_3})] .
\end{eqnarray}

Let us reproduce the main estimation of the Letter, $|\delta_D|\propto1/\sqrt{RG_Q}$, with this method. Combining Eqs. \ref{eq:delta} and \ref{eq:GG} we estimate $(|\delta_D|/\Delta)^2\propto \nu^{-1}\int dt \mathcal{P}(t) e^{-\Delta|t|}$. In the course of its propagation, an electron covers the region whose spatial dimensions are defined by the dwell time $t_{\rm dw}\simeq \Delta^{-1}$. The $\mathcal{P}(t_{\rm dw})$ is estimated as inverse volume $V$ of the region. With this we can estimate $(|\delta_D|/\Delta)^2\sim t_{\rm dw}\delta_s$, $\delta_s=(\nu V)^{-1}$. If we now compare this with the Thouless estimation \cite{thouless:pr74} of the conductance of such region, $G_{\rm Th}\simeq G_Q(\delta_s t_{\rm dw})^{-1}$, we reproduce $(|\delta_D|/\Delta)^2\propto R_{\rm Th}G_Q$. 

Let us specify the concrete setup. It comprises the semiconducting nanowire covered by three superconducting leads, see Fig.~\ref{fig:F_L_xi} (a); such devices have been recently fabricated \cite{girit:aps19}. The middle lead is a film of thickness $d$ and width $L$. If $L\lesssim\xi_0$, the ABS wave functions overlap strongly. We assume diffusive transport in the lead, which is characterized by the resistance per square $R_{\Box}$. We also assume that the interface between the nanowire and the metal is sufficiently transparent, so that the electrons escape the nanowire to metal at the distances $\ll \xi_0$.

The semiclassical propagator in the film obeys diffusion equation 
\begin{equation}
\left(\frac{\partial}{\partial t}-D\nabla^2\right)\mathcal{P}({\bm r},t)=\frac{1}{d}\delta(t)\delta({\bm r}-{\bm r}_A).
\end{equation}
This diffusion equation is subject to boundary conditions of zero probability flow across all boundaries. One satisfies these boundary conditions introducing infinite number of imaginary sources, spaced with $2L$. The propagator we obtain is
\begin{equation}
\mathcal{P}({\bm r}_A,{\bm r}_B,t)=\frac{1}{dL}\sqrt{\frac{1}{\pi D|t|}}\sum_{n=-\infty}^\infty(-1)^ne^{-D\frac{\pi^2}{L^2}n^2|t|}.
\end{equation} 
With this we compute $|\delta_D|^2$ using Eqs. \ref{eq:delta} and \ref{eq:GG} to obtain
\begin{eqnarray}
\frac{|\delta_D|^2}{\Delta^2}=\frac{\pi}{2}MG_QR_{\rm eff}F\left(\frac{L}{\xi(E_0)}\right),\ \ R_{\rm eff}=R_\Box\frac{\xi(E_0)}{L},\ \ \ 
\end{eqnarray}
with $R_{\rm eff}$ being the effective resistance of the part of the lead covered by ABS wave functions, the dimensionless $F(z)=4z/\pi\sum_{n=0}^\infty K_0((2n+1)z)$, $K_0$ being modified Bessel function of the second kind, $F(0)=1$, incorporates information of the decay of ABS wave functions at the scale of $\xi(E)=\xi_0(1-E^2/\Delta^2)^{-1/4}$, that is the energy-dependent correlation length. The prefactor $M\simeq 1$ incorporates information about transmissions of the junctions $M=2 |t_A|^2|t_B|^2[2\cos(\chi_A-\chi_B)E^2+(2+\cos(\chi_A+\chi_B)-\cos(\chi_A-\chi_B))\Delta^2-2E\Delta(\cos{\chi_A}+\cos{\chi_B})]/[\Delta^2(|t_A|^2+|t_A'|^2)(|t_B|^2+|t_B'|^2)]$, where $\chi_A$ and $\chi_B$ are the phases of the eigenvectors $|A\rangle$ and $|B\rangle$, respectively, with $e^{i\chi_A}=[|t_A'|^{2}e^{i\varphi_1}+|t_A|^2]\Delta/(E[|t_A|^2+|t_A'|^2])$ and analogously for $\chi_B$. Here, $\varphi_1$ and $\varphi_2$ denote phase differences with respect to $\varphi_3$ and we set $E_{{\rm imp},A,B}=0$. $M \rightarrow 1$, in the limit $E_0\rightarrow 0$, this requires $T_{A,B}\rightarrow 1$.
Thus in the limiting cases we have 
\begin{equation}
\frac{|\delta_D|^2}{\Delta^2}=\begin{cases}\frac{\pi}{2}MG_QR_{\rm eff}, \  L\rightarrow 0,\\
MG_QR_{\rm eff}\sqrt{\frac{2\pi L}{\xi(E_0)}}e^{-\frac{L}{\xi(E_0)}}, \  L\rightarrow \infty,
\end{cases} 
\end{equation}
We plot the normalized energy splitting $|\delta_D|/\delta_D(L=0)$ versus $L$ in Fig.~\ref{fig:F_L_xi} (b), $\delta_D(L=0)=\Delta\sqrt{\pi MG_QR_{\rm eff}/2}$. 

For the experimentally relevant parameters $\Delta=200\ \mu$eV, $\xi_0=96$ nm \cite{vanwoerkom:natphys17}, $R_\Box=1.43\ \Omega$ \cite{vanwoerkom:natphys15}, $L=50$ nm, $T_A=T_B=1$, we find the crossing point at $\varphi_2=2.36$, $\varphi_1=3.93$, and $E_0=76.54\ \mu$eV, and obtain the splitting $2|\delta_D|=11.26\ \mu$eV. The value for $|\delta_D|\simeq \Delta/40$ even though $R_\Box G_Q\approx 10^{-4}$, which seems to be small.

The separation of scales between $\delta_D$ and $\Delta$ permits interesting quantum manipulation schemes for involved states. Let us describe the simplest one: quasiparticle swap between $A$ and $B$. Let us take a point in parametric space of $\varphi_{1,2,3}$, where the ABS energies are well-split (for instance, $\varphi_3=0$ in Fig. \ref{fig:Andreev_molecule} (b)) and put a quasiparticle to the state $A$. We pass the avoided crossing slowly to avoid Landau-Zener tunneling in this point (for instance, changing $\varphi_3$ from $0$ to $\pi/2$), this brings the quasiparticle to $B$. If we get back to the initial point very quickly, the quasiparticle will remain in $B$, this completes the manipulation protocol. The same swap occurs if the quasiparticle is in $B$ initially.    

\begin{figure}[tb]
\begin{center}
\includegraphics[width=\linewidth]{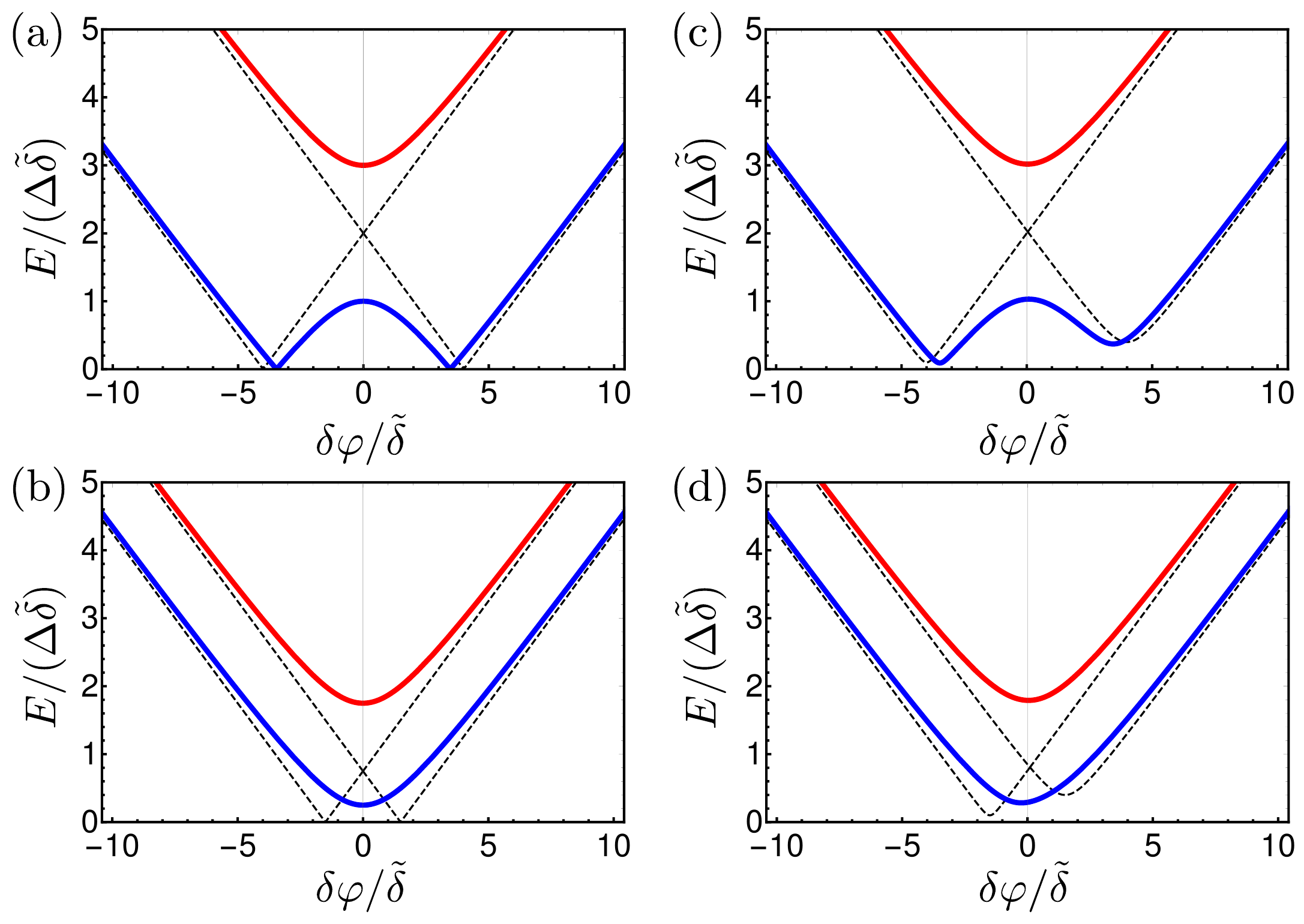}
\caption{Spectra of the system for $E\sim 0$ for $f/\tilde{\delta}=\sqrt{3/5}$, $g/\tilde{\delta}=\sqrt{2/5}$, and (a) $\Phi/\tilde{\delta}=4$, $R_A=R_B=0$; (b) $\Phi/\tilde{\delta}=1.5$, $R_A=R_B=0$; (c) $\Phi/\tilde{\delta}=4$ and $\sqrt{R_A}/\tilde{\delta}=0.1$, $\sqrt{R_B}/\tilde{\delta}=0.4$; (d) $\Phi/\tilde{\delta}=1.5$ and $\sqrt{R_A}/\tilde{\delta}=0.1$, $\sqrt{R_B}/\tilde{\delta}=0.4$. The dashed line shows the case of $f=g=0$ and for other parameters as described. The asymmetry in (c) and (d) comes from $R_A\neq R_B$.}
\label{fig:asymmetry}
\end{center}
\end{figure}

There is an interesting case, when both junctions have almost ideal transmission $T_{A,B}=1-R_{A,B}$, $R_{A,B}\ll 1$, and $\varphi_{1,2}=\pi+\delta\varphi_{1,2}$, $\delta\varphi_{1,2}\ll 1$. In this case the crossing occurs at $E_0\ll\Delta$, which can also be comparable with $|\delta_D|$. The perturbation theory does not work here, but we can describe the situation with the following $4\times4$ effective Hamiltonian:
\begin{equation}
H_{\rm eff}=\Delta\begin{pmatrix}0 & h_A & g & f\\
h_A^* & 0 & f & -g\\
g & f & 0 & h_B\\
f & -g & h_B^* & 0\end{pmatrix},
\end{equation}
where $h_{A,B}=\sqrt{R_{A,B}}+i\delta\varphi_{1,2}$. 
The terms $f$ and $g$ come from $\Sigma_{AB,BA}$, $\langle f^2\rangle=\langle g^2\rangle=16\pi^3\nu^2|t_A|^2|t_B|^2G_Q R_{\rm eff}F(L/\xi(E_0))$. In the limit $R_{A,B}=0$ the ABS energies are given by 
\begin{equation}
E=\sqrt{\tilde{\delta}^2+\frac{(\delta\varphi)^2}{4}+\frac{\Phi^2}{4}\pm\Phi\sqrt{\left[\tilde{\delta}^2+\frac{(\delta\varphi)^2}{4}\right]}},
\end{equation}
where $\tilde{\delta}=\sqrt{f^2+g^2}$, $\delta\varphi_{1,2}=\Phi/2\pm\delta\varphi/2$. Interestingly, if $|\Phi|<2\tilde{\delta}$ the energies never cross zero, while there are two symmetric zero-energy crossings if $|\Phi|>2\tilde{\delta}$, Fig.~\ref{fig:asymmetry} (a), (b). In this approximation two ABS energies are precisely degenerate at $\Phi=0$, this degeneracy is lifted upon increasing energy. At finite $R_{A,B}$ the zero-energy crossings are avoided, Fig.~\ref{fig:asymmetry} (c), (d).

Before we conclude let us mention that the fact that the energy splitting is fine makes relevant a set of topics to research that we list here. For semiconducting nanowires the electron escape length can be $\gtrsim L$, this confines the overlap region to the nanowire and greatly enhance $\delta_D$. The spin-orbit splitting \cite{yokoyama:epl14} of ABS that is usually negligible can become relevant for small $\delta_D$. Many-body effects shall provide small energy differences for doublet and singlet quasiparticle states in ABS \cite{tosi:prx19}. Interestingly, the hybridization of degenerate singlet states in the setup under investigation can also occur without direct overlap of ABS wave functions, that is at $L\gg\xi_0$.   
 
To conclude, we have investigated the energy splitting of overlapping ABS in ``Andreev molecule'' setup and have found it small in comparison with $\Delta$, since it is related to mesoscopic fluctuations. This opens up possibilities for quantum manipulation and application as mentioned in Ref. \cite{pillet:arxiv18}. The smallness of the energy splitting makes relevant the research of a variety of fine and non-trivial effects on ABS spectrum.

\begin{acknowledgments} 
We acknowledge illuminating discussions with A. Geresdi, H. Pothier, and especially with \c{C}. Girit and members of his team. This project has received funding from the European
Research Council (ERC) under the European Union's
Horizon 2020 research and innovation programme (grant
agreement No. 694272).
\end{acknowledgments}

\end{document}